\documentclass[12pt,a4paper]{article}

\usepackage{a4}

\usepackage{amsmath}
\usepackage{amsfonts}
\usepackage{amssymb}
\usepackage{multicol}  

\newcommand{\nc}{\newcommand}
\nc{\beq}{\begin{equation}}
\nc{\eeq}{\end{equation}}
\nc{\bea}{\begin{eqnarray}}
\nc{\eea}{\end{eqnarray}}
\def\ov{\overline}

\newcommand{\drawsquare}[2]{\hbox{%
\rule{#2pt}{#1pt}\hskip-#2pt
\rule{#1pt}{#2pt}\hskip-#1pt
\rule[#1pt]{#1pt}{#2pt}}\rule[#1pt]{#2pt}{#2pt}\hskip-#2pt
\rule{#2pt}{#1pt}}
\newcommand{\fund}{\raisebox{-.5pt}{\drawsquare{6.5}{0.4}}}
\newcommand{\Ysymm}{\raisebox{-.5pt}{\drawsquare{6.5}{0.4}}\hskip-0.4pt%
         \raisebox{-.5pt}{\drawsquare{6.5}{0.4}}}
\newcommand{\Yasymm}{\raisebox{-3.5pt}{\drawsquare{6.5}{0.4}}\hskip-6.9pt%
        \raisebox{3pt}{\drawsquare{6.5}{0.4}}}
\newcommand{\antifund}{\overline{\fund}}

\usepackage{epsfig}

\def\ov{\overline}

\begin{document}

\vspace*{-1.5cm}
\begin{flushright}
  {\small
 SU-ITP-09/23\\
 SLAC-PUB-13655
  }
\end{flushright}

\vspace{2cm}
\begin{center}
  {\LARGE
  Retrofitting and the $\mu$ Problem
  }
\end{center}

\vspace{0.75cm}
\begin{center}
 Daniel Green and Timo Weigand
\end{center}

\vspace{0.1cm}
\begin{center} 
\emph{SLAC and Department of Physics, Stanford University, \\ CA 94305-4060, USA }  \\

\vspace{0.3cm}

\tt{drgreen@stanford.edu, timo@slac.stanford.edu} 
\vspace{0.5cm} 
\end{center}

\vspace{1cm}


\begin{abstract}
One of the challenges of supersymmetry (SUSY) breaking and mediation is generating a $\mu$ term consistent with the requirements of electro-weak symmetry breaking.  The most common approach to the problem is to generate the $\mu$ term through a SUSY breaking F-term.  Often these models produce unacceptably large $B\mu$ terms as a result.  We will present an alternate approach, where the $\mu$ term is generated directly by non-perturtative effects.  The same non-perturbative effect will also retrofit the model of SUSY breaking in such a way that $\mu$ is at the same scale as masses of the Standard Model superpartners.  Because the $\mu$ term is not directly generated by SUSY breaking effects, there is no associated $B\mu$ problem.  These results are demonstrated in a toy model where a stringy instanton generates $\mu$.  
\end{abstract}

\clearpage


\section{Introduction}

Electro-weak symmetry breaking in the Minimal Supersymmetric Standard Model (MSSM) requires a Higgs sector with both supersymmetry breaking masses of the  form $m_{h_u}^2 h^{\dag}_u h_u $, $m_{h_d}^2 h^{\dag}_d h_d$ and $B\mu h_u h_d$ and a supersymmetric mass ($\mu \int d^2 \theta H_u H_d $).  The presence of a stable vacuum that spontaneously breaks electro-weak symmetry further requires that $\mu^2 \sim B\mu \sim m_{h}^2$.  Because $\mu$ does not break supersymmetry, there is no reason a priori that $\mu$ should be of the same scale as the supersymmetry breaking terms in the MSSM.  We will refer to this as the $\mu$ problem.

The $\mu$ term does break a Peccei-Quinn (PQ) symmetry under which both $H_u$ and $H_d$ have charge $+1$ and quarks and leptons have charge $-1/2$.  The standard approach to the $\mu$ problem is to assume that the physics of mediation is the origin of PQ breaking and thus ties the scale of SUSY breaking in the MSSM to the scale of PQ breaking.  Specifically, the $\mu$ term is generated from a K\"ahler potential term by integrating out the F-term that breaks SUSY.  

In gravity mediation, this approach can be effective via the Giudice-Masiero mechanism \cite{Giudice:1988yz}.  
Both $\mu$ and $B \mu$ are generated by tree-level effective couplings to the supersymmetry breaking field $S$ in the K\"ahler potential
\bea
K \supset    \frac{S^{\dagger}}{M_{Pl.}} H_u H_d + \frac{ S \, S^{\dagger}}{M^2_{Pl.}} H_u H_d   \quad + h.c.
\eea
This naturally leads to $\mu^2 \sim B\mu$ once one integrates out the F-term $F_S \neq 0$.

In models of gauge mediation \cite{Dine:1981za,Dimopoulos:1981au, Nappi:1982hm,AlvarezGaume:1981wy,Dine:1995ag}, reviewed e.g. in \cite{Giudice:1998bp}, the $\mu$ problem is more severe \cite{Dvali:1996cu}. Specifically, the same physics that generates the $\mu$ term will also generate a $B\mu$ term.  In the simplest models, both terms are generated by 1-loop diagrams such that $\mu^2 \ll B \mu$ due to the automatic appearance of a relative 1-loop suppression factor.  This can be evaded by introducing more elaborate messenger sectors \cite{Dvali:1996cu,Giudice:2007ca, Komargodski:2008ax} or strong dynamics \cite{Murayama:2007ge,Roy:2007nz, Ibe:2007km, Komargodski:2008ax}.

We propose an alternate approach to the $\mu$ problem, where one ties the scale of $\mu$ to the \emph{scale} of supersymmetry breaking, but not to supersymmetry breaking itself.  The $\mu$ term will be generated directly in the superpotential and thus we will not generate a $B\mu$ term simultaneously.  If one is agnostic about the origin of supersymmetry breaking, it is unclear how one would do this.  However, if one demands that supersymmetry is broken dynamically \cite{Witten:1981nf} then there may be natural mechanisms that generate $\mu$.  

We will focus on the use of retrofitting \cite{Dine:2006gm} to explain the small scale of SUSY breaking.  For the purpose of this paper, retrofitting is a mechanism by which non-perturbative effects generate the mass parameters in the supersymmetry breaking sector.  The supersymmetry breaking sector may then be as simple as a Fayet, Polonyi or O'Raifeartaigh model, with mass parameters generated by gaugino condensation \cite{Dine:2006gm} or instantons \cite{Aharony:2007db}.

The approach taken in this paper will be to generate $\mu$ by the \emph{same} non-perturbative effect that retrofits the SUSY breaking sector.  Our goal will be to produce a $\mu$ term that arises at the same scale as the soft masses in the Standard Model.  We will show how the order of magnitudes can be made to match in models with calculable non-perturbative effects.  We will focus on the use of stringy instantons, mostly for calculability of the instanton effect although one could construct purely field theoretic models with the same features.  The generation of a $\mu$ term by stringy instantons has been suggested before in \cite{Blumenhagen:2006xt,Ibanez:2006da,Buican:2006sn} with concrete realisations including \cite{Ibanez:2007tu,Ibanez:2008my,Cvetic:2009yh}.  What is new about our approach is the correlation between the scale of $\mu$ and the scale of supersymmetry breaking.

The organization of the paper is as follows:  In section \ref{sec_Retro}, we will explain the basic idea of retrofitting and $\mu$ generation.  We will also elaborate on some of the basic obstacles that can arise in specific models.  In section \ref{sec_quiver}, we will present a particular model based on a supersymmetry breaking quiver of ISS type.
Retrofitting is performed by a stringy instanton that generates supersymmetry breaking masses for the quark flavors at the same scale as the $\mu$ term.  To illustrate possible embeddings into string theory, we have constructed a toy model, based on the geometries of \cite{Blumenhagen:2008zz}, which realizes the main aspects of this quiver and the instanton sector. Technical details about the string construction are left to an appendix.
In section \ref{sec_U(1)}, we will discuss the use of anomalous $U(1)$s to prevent tree level mass terms, focusing on how this affects the couplings to the SUSY breaking sector.  We will then discuss some open issues.  

\section{Retrofitting and $\mu$}
\label{sec_Retro}

\subsection{The general idea}

As a concrete example of retrofitting, let us consider the model from \cite{Dine:2006gm}.  The dynamics of this model are non-trivial and will serve only as motivation for our own model building in later sections.  We will restate their results and refer the reader to \cite{Dine:2006gm} for details.  The superpotential is given by
\beq
W = Z_1 \frac{\phi^3}{3 M_*} + Z_2 (\lambda \frac{\phi^2}{2}[1+ \lambda_1 \frac{Z_2}{M_*}]-\lambda \frac{\sigma^2}{2} + \frac{\phi \eta \tilde{\eta}}{M_*}) + \lambda \phi \eta \tilde{\eta} + \lambda_2 \frac{(\eta \tilde{\eta})^2}{M_*},
\eeq
where $Z_1$, $Z_2$ and $\phi$ are hidden sector fields, $\eta$ and $\tilde{\eta}$ are messengers embedded in a ${\bf 5}+ {\bf \overline{5}}$ representation of $SU(5)$ and $M_*$ is some UV cutoff. The $\sigma^2$ term in the superpotential leads to SUSY breaking.

In \cite{Dine:2006gm}, the scale of SUSY breaking was generated by gaugino condensation.  Specifically, we couple an $SU(2)$ to $Z_2$ via $\int d^2 \theta W_{\alpha} W^{\alpha} \frac{\lambda Z_2}{M_*}$.  This will dynamically generate $\sigma^2 \simeq \Lambda^3 / M_*$.  In the above model, the SUSY breaking scale of the MSSM is given by
\beq
\tilde{m} \sim \frac{g^2}{16 \pi^2} \frac{\sigma^4}{M_*^3 \lambda^2} \sim \frac{g^2}{16 \pi^2} \frac{\Lambda^6}{M_*^5 \lambda^2}.
\eeq

As a simple illustration of our idea to address the $\mu$ problem, consider adding to the theory a coupling between the Higgs sector and the $SU(2)$ sector by
\beq
S \supset \int d^4 x \, d^2 \theta \, (W_\alpha W^\alpha)^2 \, \frac{\kappa \, H_u H_d}{M_*^{5}}.
\eeq
Such a coupling will generate a $\mu$ term of the order $\mu \sim \tilde{m} \sim \Lambda^6 / M_*^5$.  Of course, one would have to explain why such a highly irrelevant operator was the leading effect.  Nevertheless, it serves as a simple illustration of how one can generate $\mu$ without coupling the Higgs sector  to the messengers or SUSY breaking directly.

Furthermore, one should note that no $B\mu$ term will be generated by this coupling.  As a result, at the scale of SUSY breaking $B \mu \sim 0$.  $B \mu$ and $m_h^2$ will be generated through renormalization group (RG) flow in the MSSM.  This will lead to a consistent Higgs sector provided the gaugino masses ($m_{\tilde{g}}$) are at the weak scale and $m_{h}^2 \lesssim m_{\tilde{g}}^2$ at the SUSY breaking scale.  Thus, electroweak symmetry breaking will require 
\beq
\label{constraint}
\mu \sim m_{\tilde{g}}.
\eeq
A viable model should satisfy this condition after including all order one factors.   These factors are difficult to calculate when considering non-perturbative effects and/or UV complete models.  We will discuss specific factors as they arise, but the end result will contain a theoretical error competitive with factors of $16 \pi^2$.  It is important that some of these factors are functionally independent for $\mu$ and $m_{\tilde{g}}$ such that there is no in principle obstacle to finding viable models.  
As reviewed in the introduction, this is in contrast to the generation of $\mu$ by Giudice-Masiero type 1-loop K\"ahler potential couplings in gauge mediation, where the $16 \pi^2$ discrepancy to the $B \mu$ term is generic \cite{Dvali:1996cu} and avoidable only with extra effort  \cite{Dvali:1996cu,Giudice:2007ca, Komargodski:2008ax, Murayama:2007ge,Roy:2007nz, Ibe:2007km}.

Finally, note that
one may also consider cases where $B \mu$ is generated by messenger couplings to the Higgs, leading to Giudice-Masiero type couplings at one loop. Since the $\mu$ term is generated independently at the correct scale by non-perturbative effects, for appropriate values of the couplings a viable $B \mu$ term is achievable. In this case, the additional contribution to the $\mu$ term produced by the same physics responsible for the $B \mu$ term is subleading.

As a starting point, we will look to satisfy (\ref{constraint}) at the order of magnitude level.  For the time being, we will assume that couplings like $\kappa$ are order one, but we will return to discuss how couplings are generated in section \ref{sec_U(1)}.

\subsection{Stringy Instantons}

For the rest of the paper, we will focus on the use of (stringy) instantons to generate the necessary non-perturbative effects (see \cite{Blumenhagen:2009qh} for a recent review).   The advantage of this approach is that conditions for generating non-perturbative superpotential terms can be determined geometrically. 

To demonstrate the general idea we will think of field theories that arise on D-branes wrapping cycles in some \emph{local} geometry. This effectively reduces the problem to supersymmetry breaking in globally supersymmetric field theories. 
The successful realisation of supersymmetry breaking in compact string constructions is a much more intricate problem which requires full control of the moduli stabilisation sector in supergravity and is beyond the scope of this article. We will comment on associated problems in section \ref{moduli}.

 The gauge groups arise from the number of D-branes on a given cycle.  Bi-fundamental matter ($\Phi_j$) are localised at the intersections of various cycles.  A Euclidean brane wrapping a cycle of volume ${\rm Vol}$ in units of the string length gives rise to non-perturbative effects suppressed by $e^{-S_0} = e^{-\frac{\rm Vol}{g_s}}$.  Our goal will be to use a single instanton to generate both the $\mu$ term and a mass term in the SUSY breaking sector.

In order to compute the contributions to the superpotential, one must determine the zero modes of the instanton.   The universal zero modes arise from the breaking of translation invariance and SUSY in the presence of the instanton.  For the Euclidean brane to give a superpotential contribution, integrating over these zero modes should introduce $\int d^4 x \, d^2 \theta$.  In addition, there are charged fermionic zero modes ($\alpha_i$) due to massless strings that stretch between the Euclidean brane and D-branes on other cycles \cite{Ganor:1996pe, Blumenhagen:2006xt,Ibanez:2006da,Florea:2006si}.   The number of such modes can be determined in the same way that the number of bi-fundamental fields is determined from the intersection of branes.

The Euclidean brane contributes to the action a term
\beq
S \supset \int M_s^3 \, d^4x \, d^2 \theta   \int  \prod_i d\alpha_i  \, e^{-S_0 - S_{int}(\alpha_i, \Phi_j)},
\eeq
where $M_s$ denotes the string scale. Here $S_0 = \frac{\rm Vol}{g_s}$ with $g_s$ the string coupling and   $S_{int}(\alpha_i, \Phi_j)$ denotes the couplings between the charged instanton zero modes and the open string modes in the D-brane sector. This interaction piece can be determined by the same techniques that determine the couplings between the matter fields \cite{Cvetic:2007ku,Kachru:2008wt}.  Integrating out the fermionic zero modes $\alpha_i$ will generate superpotential terms involving the matter fields $\Phi_j$.

In \cite{Aharony:2007db}, it was shown how these contributions can be used to engineer simple local models of SUSY breaking  (see  \cite{Cvetic:2007qj,Buican:2008qe,Cvetic:2008mh,Heckman:2008es,Marsano:2008jq, Heckman:2008qt} for concrete realisations and applications).
For example, one can construct the Polonyi model if there are only two fermionic zero modes, $\alpha$ and $\beta$, that couple in the instanton action to a field $\Phi$ via $S_{int} = c_{F} M_s^{-1}\, \alpha \Phi \beta$.  The coupling  $c_{F}$ is a function of the open and closed string moduli and therefore model dependent.   Integrating out $\alpha$ and $\beta$ we get a superpotential for $\Phi$ of the form
\beq
W = \sigma^2 \Phi \simeq c_{F} \, e^{-S_0} \, M_s^2 \, \Phi.
\eeq
Here we are  neglecting an additional ${\cal O}(1)$ factor due to the 1-loop Pfaffian of the instanton path integral.
In models of this type, one generates an F-term of the form $F_{\Phi} = \sigma^2 \simeq c_F \, e^{-S_0} M_s^2$.

What we need is that $\mu$ is of  the scale of SUSY breaking in the MSSM.  On dimensional grounds, any simple model of mediation should generate soft masses of the form
\beq
\label{gaugino}
m_{\tilde{g}} \sim \frac{\lambda(\phi,r,m,M,\ldots)}{16 \pi^2} \frac{F}{M},
\eeq
where $\lambda(\phi,F,M,\ldots)$ is some dimensionless quantity that may depend on the dimensionful and dimensionless parameters of the theory, and $M$ is some mass scale associated with the messengers. 
However, if we generate $\mu$ by the same instanton, then we get
\beq
\label{cmu}
W \supset \mu \, H_u \, H_d \simeq c_{\mu} \, M_s \, e^{-S_0} \, H_u \, H_d,
\eeq
with $c_{\mu}$ denoting the coupling in $S_{int} \supset c_{\mu} \, M_s^{-2} \, \alpha \, H_u\,  H_d \, \beta$.
For $\mu \simeq m_{\tilde g}$ we need 
\beq
\label{MPol}
M  \simeq \frac{\lambda}{16 \pi^2} \frac{c_{F}}{c_{\mu}} \, M_s.
\eeq
Note that since in typical string setups $c_F$ and $c_{\mu}$ can easily differ by factors competing with $16 \pi^2$, this relation might result in string scale messengers, depending on the details of the model.
A related potential worry is that domination of gauge mediation over gravity mediation requires $ \frac{\lambda}{16 \pi^2} \frac{F}{M} < \frac{F}{M_{Pl.}}$, so typically $M \leq 10^{-3} M_{Pl.}$.
This can be read as a constraint of the model dependent properties of the dimensionless coupling constants appearing in the setup.
In this sense models of SUSY breaking where a $(mass)^2$ controls SUSY breaking will generically be more involved.  A possible resolution to this problem is to generate the couplings in the messenger sector non-perturbatively as well, as we will discuss in section  \ref{sec_U(1)}.

For the above reasons, it will be technically simpler to focus on models where a coupling of mass dimensions one is generated by the instanton.  A simple example of this type is the Fayet model.  In this model, we have a $U(1)$ gauge field that couples to two chiral superfields $\Phi_{\pm}$ with opposite charges.  We wish to generate a superpotential
\beq
W = m \,  \Phi_{+} \,  \Phi_{-} = c_m \, M_s \, e^{-S_0} \Phi_{+} \Phi_{-}. 
\eeq
SUSY is unbroken if both $\phi_{\pm} = 0$.  If the theory has a non-zero Fayet-Iliopoulos (FI) parameter $r$, then the D-term requires that
\beq
|\phi_+|^2- |\phi_-|^2 = r.
\eeq
For $\sqrt{r} \gg m$, this leads to F-term breaking where $|\phi_{+}|^2 \sim r$ and $F_{\Phi_{-}} \sim m \sqrt{r}$.

One could construct such a model where $m$ is generated by a stringy instanton, as in \cite{Aharony:2007db}, and generate a $\mu$ term of the form (\ref{cmu}) from the same instanton.  As a result we will get $\mu = \frac{c_{\mu}}{c_m} \, m$.  
Then $\mu \sim m_{\tilde{g}}$ requires
\beq
\label{M_Fayet}
M \sim \frac{\lambda}{16 \pi^2} \frac{c_m}{c_{\mu} } \sqrt{r}.
\eeq

On the one hand, the scale of $M$ is now free in principle (as long as $M \gg \mu$) and not directly determined by the scale of supersymmetry breaking. This resolves potential conflict with the requirement of dominance of gauge versus gravity mediation.
Still, the relationship (\ref{M_Fayet})  is not generic and would have to be ensured by model building in the field theory or by some special features of the geometry (i.e. from the UV completion).  For the FI parameter, engineering such a relationship seems difficult. In the next section, we will see one way such a relationship can occur naturally.  

None of the problems in this section seem insurmountable, but do require more input than just a single field model of SUSY breaking.  Having a natural mechanism for generating $\mu$ in such simple models makes this worth further study.  We will now move on to more complicated models of SUSY breaking where we know of simple solutions to these basic obstacles.

\section{A Quiver for $\mu$}
\label{sec_quiver}

\subsection{The quiver}

As we discussed in the previous section, if we wish to generate both $\mu$ and SUSY breaking from a single instanton, it is most convenient to have the $\mu$ term and the supersymmetry breaking operator of the same mass dimension.  As in the Fayet model, we would then expect $F \sim \mu M$ where $M \gg \mu$ is some mass scale in the SUSY breaking sector.

A model where SUSY breaking is F-term dominated and is the product of two mass scales is the construction of ISS \cite{Intriligator:2006dd}.  We will consider $SU(N_c)$ gauge theories with $N_f$ flavors ($q_i, \tilde{q}_i$) such that $3 N_c / 2 > N_f > N_c$.  If all the $N_f$ flavors have  a mass in the superpotential 
\beq
W \supset m \, q_i \, \tilde{q}_{i},
\eeq 
there is a metastable vacuum which breaks SUSY with 
\beq
F \simeq m \Lambda,
\eeq 
where $\Lambda$ is the dynamical scale of the $SU(N_c)$ gauge theory.  Like in the Fayet model, we will generate both a $\mu$ term for $H_u H_d$ as well as the mass $m$ for the flavors by one instanton.

In order to find a string construction of the model, including the appropriate instanton effect, it will be convenient to write the model as a quiver diagram.  In the interest of simplicity, our ``Standard Model" will be an $SU(5)$ node with vector-like matter in the $({\bf 5} + {\bf \ov 5} )$ representation yielding our $H_u$ and $H_d$.  We also want to have an $SU(N_c)$ node with $N_f$ flavors for SUSY breaking.  We will communicate SUSY breaking to the Standard Model by a set of vector-like messengers ($\Psi, \tilde{\Psi}$) in the $({\bf 5}+{\bf \overline{5}})$ representation of $SU(5)$ and $({\bf N_c} + {\bf \overline{N}_c})$ representation of $SU(N_c)$.

 Typically in stringy models of ISS, one uses a ``flavor brane" to engineer $N_f$ flavors.  This would mean that the quarks would be bi-fundamentals $(\bf 5, N_f )$  of $SU(5)$ and an $SU(N_f)$ gauge group realised on a stack of $N_f$ branes.  This will not work for us, as the fermionic instanton zero modes would have to be charged under the $SU(N_f)$ or the $SU(N_c)$ gauge groups.  Integrating out these zero modes would then produce a term of the form $\det(q\tilde{q})$ rather than $m q_i \tilde{q}_i$.  Therefore, we will require that the $N_f$ flavor symmetry is not gauged.  In our string construction, this can be achieved by introducing a single brane with gauge group $U(1)$ and $N_f$ copies of messengers $(\Psi_,  \tilde \Psi)$ in the bi-fundamental representation $({\bf 5}, -1)$ and  $({\bf \ov 5},1)$.  The $U(1)$ gauge group is massive via the St\"uckelberg mechanism and remains only as a global perturbative symmetry.

\begin{figure}[t] 
	   \centering
	   \includegraphics[width=3in]{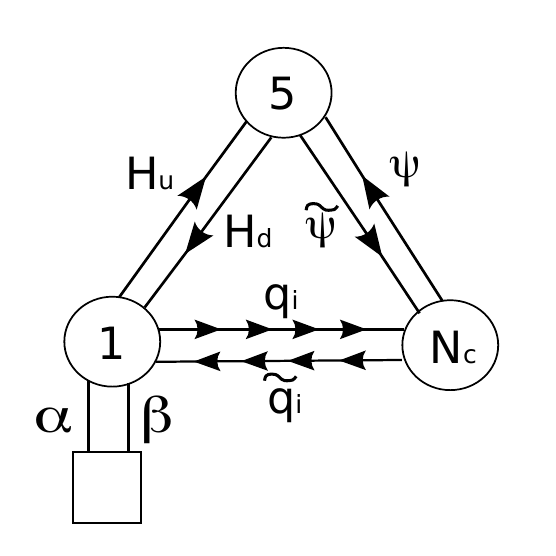} 
	   \caption{A quiver diagram for the theory described in section \ref{sec_quiver}.  The numbers give the rank of $U(N)$ gauge groups and the arrows represent bifundamental matter.  The square node is the stringy instanton that gives rise to the $\mu$ term and the mass for the flavors, where $\alpha$ and $\beta$ are the charged zero modes of the instanton. }
	   \label{fig:a}
	\end{figure}

Likewise the Higgs and flavor quarks arise as bi-fundamental fields in the $5-1$ and $3-1$ sector.
There are in principle two different possible  assignments of anomalous $U(1)$ charges to the Higgs and flavor quarks: Either $H_u$, $H_d$ form proper vector-like pairs in the representation
$({\bf 5},-1)$, $({\bf \ov 5},1)$ (and similarly for $q_i, \tilde q_i$).  
To engineer  the correct instanton effect,  the instanton should only have {\it two} charged zero modes, $\alpha$ and $\beta$, of charge $-1$ and $+1$, respectively.  
In addition the instanton action must contain the terms
\beq
S_{int} = \lambda_F^{ij } \, M_s^{-2} \, \alpha \, q_i \, \tilde{q_j}\,  \beta + c_{\mu} \, M_s^{-2} \, \alpha \, H_u \, H_d \, \beta,
\eeq
where $c^{i j}_{F}$ is a rank $N_f$ matrix.  Integrating out $\alpha$ and $\beta$ will produce, up to the couplings, 
\beq
m \sim \mu \sim e^{-S_0}.
\eeq  The full quiver, including the instanton, is shown in figure \ref{fig:a}.

The disadvantage of this charge assignment is that no symmetry protects the $\mu$ term and the quark masses from being generated perturbatively, i.e. generically at the string scale. This can be remedied by assigning charges  $({\bf 5},+1)$, $({\bf \ov 5},+1)$ to $H_u, H_d$ and $({\bf N_f},+1)$, $({\bf \ov N_f},+1)$ to $q_i, \tilde q_i$ as well as $-1$ to both $\alpha$ and $\beta$,  see table \ref{tab_tod1}.
Such charges can easily be achieved in a stringy orientifold setup. The operators $H_u H_d$ and $q \tilde q$ thus carry charge $+2$ under the massive $U(1)$, which is now also anomalous and whose anomaly is cancelled by the Green-Schwarz mechanism. The charges of the instanton zero modes guarantee that the volume modulus appearing in $e^{-S_0}$ shifts appropriately so as to render the non-perturbative $\mu$ and mass term gauge invariant.

\begin{table}[htbp] 
\renewcommand{\arraystretch}{1.5} 
\begin{center} 
\begin{tabular}{|c||c|c|c|c|| c|c|c|c|c|} 
\hline 
\hline 
field & $U(5)$ & $U(N_c)$ & $U(1)$      \\ 
\hline \hline 
$q_j$ & $1$ & $\fund$ & $1$ \\ 
$\tilde{q}_j$ & $1$ & $\antifund$ & $1$ \\
$H_u$ & $\fund$ & $1$ & $1$  \\
 $H_d$ & $\antifund$ & $1$ & $1$  \\
\hline 
\end{tabular} 
\caption{Charges associated with the fields $H_u, H_d$ and $q_i, \tilde q_i$ in the quiver in figure \ref{fig:a}.} 
\label{tab_tod1}
\end{center} 
\end{table}

In the low energy theory, it is the meson field 
\beq
\tilde{M}^j_i = \Lambda^{-1} \, q_i \, \tilde{q}^j
\eeq
 which acquires a SUSY breaking F-term , $F_{\tilde M} = m \Lambda \simeq \mu \Lambda$.   The messengers will have a tree level mass,
 \beq
 W  \supset M \Psi \tilde \Psi.  
 \eeq
 In the string construction $M$ will be controlled by an open string modulus, which has to be tuned to get $M \ll M_s$.  We will couple the messengers to the mesons via $W \supset \kappa_{j}^{i} \tilde{M}^j_i \Psi \tilde{\Psi}$.   Clearly the second charge assignment, where $q_i \tilde q_i$ carries net charge $+2$ under the anomalous $U(1)$, suffers from the drawback of forbidding this coupling at the perturbative level.
Slight generalizations of this model, to be introduced in section \ref{sec_U(1)}, will overcome this deficiency.
The coupling $\kappa$ will depend on $\Lambda / M_*$, where $M_*$ is some mass scale where the coupling of the messengers to the quarks is generated.  The coupling of ISS models to a messenger sector was analyzed in \cite{Murayama:2006yf}, although there  $M_*$ is identified with the Planck scale.

Our model has many features in common with the Fayet model we discussed above.  The gaugino masses in the ``Standard Model" are given by
\beq
m_{\tilde{g}} \simeq \frac{\kappa}{16\pi^2} \frac{F}{M} \sim \frac{\kappa}{16\pi^2} \, \mu \frac{\Lambda}{ M},
\eeq
where we used the fact the instanton gives $m \sim \mu$.  To ensure that $\mu \sim m_{\tilde{g}}$, we will need $\Lambda \sim M$.  Unlike in the Fayet model, there is a straightforward way to satisfy this condition.  Since $M \gg \mu$, the relevant dynamical scale $\Lambda$ is determined after integrating out the messengers.  If the UV theory is weakly coupled at the scale $M$, we can use the 1-loop matching formula to determine the new dynamical scale ($\Lambda$) from the one determined by the string construction ($\tilde{\Lambda}$),
\beq
\label{match}
\Lambda^{b(N_f)} \sim \tilde{\Lambda}^{b(N_f + 5)} M^{b(N_f) - b(N_f + 5)},
\eeq
where $b(N_f) \simeq 3 N_c - N_f$ is the coefficient of the 1-loop beta function for the coupling of the $SU(N_c)$ gauge theory.  The exponent $b(N_f + 5)$ appears because the messengers are in the $5+ \overline{5}$ of $SU(5)$ and thus act like five heavy flavors.  This matching is only approximate and more precise matching calculations will lead to order one factors in the relationship between $\Lambda$ and $M$.

We see from (\ref{match}) that the condition that $\Lambda \sim M$ arises naturally if $b(N_f+5) \ll b(N_f)$.  The simplest way to achieve this is if the 1-loop contribution to $b(N_f+5)$ vanishes, which occurs if $N_f + 5 \sim 3 N_c$.  If this is the case, we naturally get $\Lambda \sim M$ and therefore $\mu \sim m_{\tilde{g}}$.  Let us consider the case where 
\bea
N_c = 3, \quad\quad  N_f = 4.
\eea
Clearly, $N_f + 5  = 9 = 3 N_c$, so at 1-loop $b(N_f+5) \sim 0$.  The 2-loop contribution leads to a Landau pole, but we will UV complete with string theory below that scale.  Of course, because this is a purely field theoretic mechanism, we will need $M \ll M_s$ so that we are well within the field theory regime.  We will see in our string construction that this requires a small amount of fine tuning.

\subsection{String Construction}

In this section, we will give a qualitative description of a string theoretic realisation of the quiver in figure \ref{fig:a}, leaving all technical details to the appendix.
The quiver shown in figure \ref{fig:a} naturally appears in the construction of \cite{Blumenhagen:2008zz}.  This construction is based on an orientifold compactification of Type IIB string theory with D7 branes carrying worldvolume flux.  We stress that even though this setup is compact, we treat the open and closed string moduli controlling the mass scales and supersymmetry breaking scales as tunable parameters of the theory. Clearly in a compact vacuum this is not justified. The stabilisation of these moduli in the present context is  a central question, both to decide if supersymmetry may be restored in moduli space and in order to estimate the dominance of gauge mediation. This important open issue is common to all fully-fledged embeddings of supersymmetry breaking gauge dynamics into string theory and not specific to our proposed solution of the $\mu$ problem. We therefore think it is useful to demonstrate how at least the basic ingredients of our proposed quiver can be engineered in string theory models.

The three different gauge groups arise from different spacetime-filling stacks of D7 branes wrapping internal four-cycles.  As can be seen in the original models of \cite{Blumenhagen:2008zz}, canceling the D7 brane charge for an $SU(5)$ GUT gives precisely the $U(1)$ and $SU(3)$ gauge groups that our model requires.\footnote{The diagonal $U(1)$ factors of the original $U(5)$ and $U(3)$ groups on the brane stacks become massive and will play no role here.}

The number of quark flavors is determined by the intersection form of the geometry and the worldvolume fluxes on the D7 branes.  For a concrete choice of cycles on which D7s are wrapped in a specific geometry, getting $N_f = 4$ puts a condition on the worldvolume flux (line bundle).

A Euclidean D3 brane wrapping another cycle not populated by any of the $SU(5)$, $U(1)$ and $SU(3)$ branes will generate both the quark masses and a $\mu$ term. Let us focus for definiteness on the second $U(1)$ charge assignment summarized in table \ref{tab_tod1}, for which this non-perturbative effect is the leading order contribution to both terms due to the global $U(1)$ symmetry. 
The Euclidean brane must not carry any worldvolume flux to contribute to $W$, and  must have non-trivial intersection with the D7 on the $U(1)$ node to allow for charged zero modes.  The requirement of precisely two such modes $\alpha, \beta $ between the $U(1)$ node and the Euclidean brane and no zero modes charged under the other gauge groups puts extra constraints on the geometry. In principle there do exist realisations of this scenario in the present setup with exactly one pair of $H_u, H_d$ fields, but due to a certain technical complication discussed in the appendix 
we discard these solutions here. The smallest number of Higgs pairs avoiding this complication turns out to be $4$ in our particular geometry. This is just an unfortunate feature of the comparatively simple geometry we are considering and we therefore believe that this model is sufficient for the purpose of illustrating the mechanism.
The explicit data of an example of such a model is given in equ. (\ref{divdata}) and (\ref{bundata}).

The messenger fields in the model acquire mass at tree-level, which means they would generically be string scale.  Similar situations occur in many D-brane constructions, where vector-like matter has a mass proportional to the separation between branes (for a discussion in the context of gauge meditation see \cite{Diaconescu:2005pc}).  By tuning the distance between the branes to be smaller than the string scale, one engineers massive particles in the low energy effective theory.  It is important that when one is tuning open string moduli to achieve masses small, one does not necessarily introduce other light fields corresponding to closed string modes.

Similarly in our model, the mass of the messengers is controlled by an open string modulus.  Thus, like the distance between branes, we can ``tune" the mass  to be below the string scale.\footnote{Again, eventually the open string moduli will have to be stabilized dynamically.}  We will construct a simple toy model to illustrate the open string moduli dependence of the messenger mass.  The $SU(3)$ D7 branes in (\ref{divdata}) can be split into two separate stacks of branes. In that limit, there are two $U(1)$ gauge fields, one on each of the two cycles after splitting, and two additional chiral fields, $\phi_{\pm}$ stretching between these cycles in both directions.  $\phi_{\pm}$ has change $\pm 1$ under the first $U(1)$ and $\mp 1$ under the second.  Each messenger only transforms under one of the $U(1)s$ such that $\phi_{-} \Psi \tilde{\Psi}$ forms a gauge invariant combination.  The D-terms for this system take the form
\beq
\label{Dterms}
(|\phi_+|^2 - |\phi_-|^2 - \zeta_1)^2 + (- |\phi_+|^2 + |\phi_-|^2 - \zeta_2)^2,
\eeq
with the field dependent FI terms $\zeta_i$ determined by the gauge flux along the two components of the divisor.
A supersymmetric configuration requires $\zeta_1 = - \zeta_2$. In this case  the D-term conditions force $\langle \Phi_-\rangle =\langle \phi_+\rangle  - \langle \phi_-\rangle  =0$, while the combination $\Phi_+= \phi_+ + \phi_-$ remains as a flat direction (modulus).  A non-zero  vev  $\langle \phi_+\rangle  = \langle \phi_-\rangle$
breaks $U(1) \times U(1) \to U(1)$ in a way consistent with recombination of the branes.  The  vev of  the modulus $\Phi_+$ gives a mass $\frac12 \langle \Phi_+ \rangle \Psi \tilde{\Psi}$  to the messengers if there exists a superpotential $W = \phi_{-} \Psi \tilde{\Psi}$.  It is this open string modulus that controls the mass of our messengers.

There are two further obstacles which generically arise in all such models when trying to make sense of the model as a consistent compact string vacuum.  First of all, we were unable to cancel the D3 tadpole while maintaining the matter content of the model (see the appendix for details).  This is can be resolved in priniciple with a more complicated orientifold.  The second and conceptually much more serious problem is moduli stabilization, see section \ref{moduli} for more comments.

\section{Massive $U(1)$s and Couplings}
\label{sec_U(1)}

\subsection{An extended quiver}

As pointed out already, it is important to explain the absence of tree level contributions to the non-perturbative mass terms for the SUSY breaking sector or the $\mu$ term.  Without further symmetries these superpotential terms are technically natural in field theory and thus one often argues they can be freely chosen and will only be corrected non-perturbatively.  However, in string models, it is rarely the case that operators allowed by gauge symmetries are absent from the superpotential.

A common way to forbid dangerous superpotential terms in string theory is through anomalous $U(1)$ charges.  If the operators of interest carry charge under some set of $U(1)$ gauge fields, then they are forbidden by the gauge symmetry.  Since the anomalous $U(1)$ becomes massive via the Green-Schwarz mechanism, these operators may be generated non-perturbatively. As reviewed previously, the instanton action is proportional to the closed string moduli that shift to cancel the anomaly.  As a result, $e^{-S_0}$ can carry the correct charge to generate the mass terms of interest  \cite{Blumenhagen:2006xt,Ibanez:2006da,Florea:2006si}.

\begin{figure}[t] 
	   \centering
	   \includegraphics[width=3in]{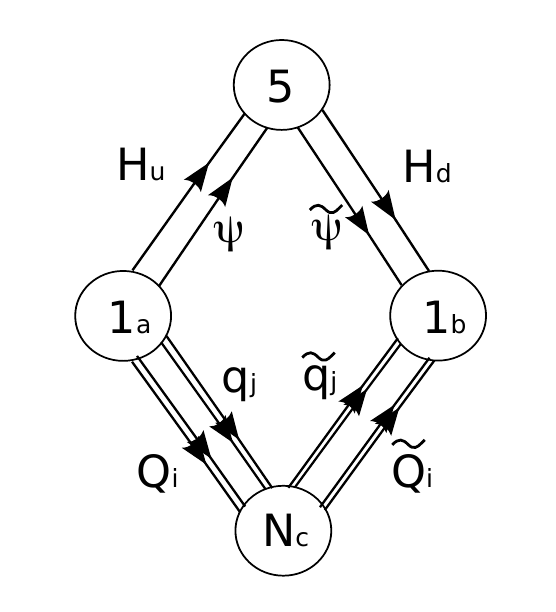} 
	   \caption{A quiver diagram for the theory described in section  \ref{sec_U(1)}.  The numbers give the rank of the $U(N)$ gauge groups and the arrows represent bifundamental matter.  The number $N_f$ of $q$ and $\tilde{q}$ fields is not determined. The number of $Q$ and $\tilde{Q}$ fields is given by $\tilde{N}$.  SUSY breaking will require that $3/2 N_c > N_f > N_c$ and solving the $\mu$ problem will require that $\tilde{N} = 3 N_c - N_f$.}
	   \label{fig:b}
	\end{figure}

If anomalous $U(1)$s are used to justify the superpotentials of our model, one must also consider the origin of the messenger mass and the coupling of SUSY breaking to the messengers.   As we saw in the previous section, if the flavors are charged but the messengers are uncharged, then both the mass term for the flavors and the coupling of the flavors to the messengers will be forbidden at tree level.

One possible way out would be to consider the case where also the messenger couplings are generated non-perturbatively.  For instance this opens the possibility of having a Polonyi model as the SUSY breaking sector without the need to require string scale messengers, see the discussion around (\ref{MPol}).  Specifically, if the coupling and the messenger mass were both generated by the same non-perturbative effect, then we could have $\lambda = e^{-S_1}$, $M \sim M_s \, e^{-S_1}$, $\mu \sim M_s \, e^{-S_2}$ and $F= M_s^2 \, e^{-S_2}$.  This would give gaugino masses at the scale
\beq
m_{\tilde{g}} \sim \frac{\lambda F}{M} \sim M_s \, e^{-S_2} \sim \mu.
\eeq
Unfortunately, we were unable to realize such a model where all the operators carry charge.  Such a model may exist, but will likely involve a more elaborate messenger sector.

\begin{table}[htbp] 
\renewcommand{\arraystretch}{1.5} 
\begin{center} 
\begin{tabular}{|c||c|c|c|c|| c|c|c|c|c|} 
\hline 
\hline 
field & $U(5)$ & $U(N_c)$ & $U(1)_a$ & $U(1)_b$  &field & $U(5)$ & $U(N_c)$ & $U(1)_a$ & $U(1)_b$   \\ 
\hline \hline 
$Q_i$ & $1$ & $\fund$ & $-1$ & $0$   &    $\Psi$ & $\fund$ & $1$ & $-1$ & $0$    \\
$\tilde{Q}_i$ & $1$ & $\antifund$ & $0$ & $1$   &$\tilde{\Psi}$ & $\antifund$ & $1$ & $0$ & $1$  \\
$q_j$ & $1$ & $\fund$ & $1$ & $0$  & $H_u$ & $\fund$ & $1$ & $1$ & $0$ \\
$\tilde{q}_j$ & $1$ & $\antifund$ & $0$ & $-1$ & $H_d$ & $\antifund$ & $1$ & $0$ & $-1$ \\
\hline 
\end{tabular} 
\caption{Charges associated with the fields in the quiver in figure \ref{fig:b}.} 
\label{tab_tod2} 
\end{center} 
\end{table}

The quiver of section \ref{sec_quiver} can be modified in order to allow for order one couplings but forbid tree level mass parameters.  This modified quiver is shown in figure \ref{fig:b}.  Both $U(1)$ nodes will have anomalies that are cancelled by the Green-Schwarz mechanism.  We will include $N_f$ fields $q_j$ and $\tilde{N}$ fields $Q_i$.  For the purpose of SUSY breaking, the rank of gauge group $N_c$ is such that $3/2 N_c > N_f > N_c$.  The charge assignments for the fields are displayed in table \ref{tab_tod2}.  For these charges, the  possible tree level superpotential terms begin at quartic order, 
\beq
\label{W_4}
W_{tree} = \frac{\lambda^{(1)}_{ij}}{m_*} Q_i \tilde{Q}_j H_{u} H_d +\frac{\lambda^{(2)}_{ij}}{m_*} q_i \tilde{q}_j \Psi \tilde{\Psi}
+  \frac{\lambda^{(3)}}{m_*} \Psi \tilde{\Psi} H_{u} H_d + \frac{\lambda^{(4)}_{ijkl}}{m_*} Q_i \tilde{Q}_j  q_i \tilde{q}_j.
\eeq
Clearly the presence of such couplings, the rank of coupling matrices $\lambda^i$ and the mass scale $m_*$   are model dependent.
The only coupling we need for gauge mediation to work is the second one.

As before, we will use an instanton to generate both $\mu H_{u} H_{d}$ and $\mu q_j \tilde{q}_j$.  For the instanton action to produce these terms, the instanton must have a zero mode with charge $-1$ under $U(1)_a$ and another zero mode with charge $+1$ under $U(1)_b$.  This will generate $\mu \sim M_s e^{-S_1}$.  A second instanton will generate the mass $M \Psi \tilde{\Psi}$ for the messengers  and  $M Q_i \tilde{Q}_i$ for the heavy flavors.\footnote{Note that the effective mass terms for the messengers and quarks after condensation of the Higgs fields in (\ref{W_4})  will be only a negligible correction.} This requires that the instanton has a charged zero mode with charge $+1$ under $U(1)_a$ and another zero mode with charge $-1$ under $U(1)_b$.  This will generate $M \sim M_s e^{-S_2}$.  We want the messengers to be much heavier than the Higgs, so we will need $S_2 < S_1$.

The structure of the quiver is different from that of figure \ref{fig:a} in order to give the messengers $U(1)$ charge.  This is needed to allow for tree level couplings.  Fortunately, this will also allow for non-perturbative messenger masses at no additional cost.  This is not possible in models where the messengers are charged under both the Standard Model and SUSY breaking gauge groups.  In such cases, the charged zero modes must carry charge under a non-abelian group and thus we get determinants rather than mass terms in the superpotential.

The lower part of the quiver in figure \ref{fig:b} is our SUSY breaking sector, as in the previous section.  The only difference in this case is that the messengers are not the heavy flavors.  This will also give us extra freedom to choose both the gauge group and the number of flavors.  Following our matching argument of the previous section, we find the $\Lambda \sim M$ when $\tilde{N} = 3 N_c - N_f$.  Although the messengers are not directly involved in the matching condition, the fact that the heavy quark masses arise from the same instanton as the messenger mass ensures the result is the same.

The tree level couplings in (\ref{W_4}), if present, also lead to one-loop diagrams that generate $\mu$ and $B\mu$ after integrating out the messengers.  Specifically, the contribution to $B\mu$ is given by
\beq
B\mu \simeq \frac{ \kappa^2}{16 \pi^2} \frac{ \lambda^{(3)} |F|^2}{m_* M},
\eeq
where $\kappa \sim \lambda^{(2)}\Lambda / m_*$.  This should be expected as we have direct couplings of the messengers to the Higgs sector.  In the absence of our instanton contribution to $\mu$, such a model would lead to an unacceptable Higgs sector \cite{Dvali:1996cu}.  The reason is that one cannot independently set the scales of $\mu$ and $B\mu$. Since $B\mu$ is generically much larger, this gives back the usual $\mu$ problem in gauge mediation.  However, with the instanton, we now have independent control over $\mu$.  The couplings in (\ref{W_4}) are only constrained by $ B\mu \lesssim m^2_{\tilde{g}} \simeq |\frac{\kappa F}{16 \pi^2 M}|^2 $.  For given values of  $\lambda^{(2)}_{ij}$ and all the dimensionful parameters, this will lead to an upper bound on the size of the coupling $\lambda^{(3)}$.  In general, $\lambda^{(3)} \leq 10^{-2}$ is sufficient but $\lambda^{(3)} \sim \mathcal{O}(1)$ is permitted when $m_* \gg M$. Since generically $m_* = M_s$, this is easily achieved.

Unlike our previous model, we have not attempted to find a geometry where the quiver and instantons arise, but there is no obstacle of principle to doing so. Given a geometry that yields a quiver of the form given in figure \ref{fig:b}, the techniques described in the appendix can be used to engineer the correct number of bifundamentals on the different nodes, as well as instantons that will produce the mass terms that are needed for this model.

\subsection{Remarks on moduli stabilisation}
\label{moduli}

Our discussion has treated the moduli controlling the couplings of the models as free parameters and is therefore valid in the limit of a supersymmetric field theory. 
We have repeatedly alerted the reader, though, that a full realisation of supersymmetry breaking and gauge mediation in \emph{compact} string models is severely complicated by the requirement of moduli stabilisation.  The problem comes in two parts. Firstly, in the presence of any non-trivial D-brane sector it is difficult to stabilize the volume moduli of the D-branes in an appropriate regime. This problem was raised in \cite{Blumenhagen:2007sm} in the context of Type IIB orientifolds. There it was pointed out that those D-brane instantons which are usually responsible for stabilisation of the K\"ahler moduli do not generate the necessary dependence of the superpotential on all the K\"ahler moduli due to extra zero modes related to the presence of the D-branes. This challenge calls for a solution in most attempts to realize realistic phenomenology and is per se  independent of the engineering of a supersymmetry breaking gauge sector. More specifically applied to scenarios of supersymmetry breaking  this means that e.g. for simple realizations of Polonyi type models by D-brane instantons as in  \cite{Aharony:2007db,Cvetic:2007qj,Buican:2008qe,Cvetic:2008mh,Heckman:2008es,Marsano:2008jq,Heckman:2008qt}   it remains to show how the moduli controlling the volume of the instanton cycle are stabilised in a regime where the F-term is non-vanishing. Aspects of this problem have been analyzed for the simplest examples in \cite{Dudas:2008qf}. In the context of our quiver, the volume instanton generating the $\mu$ term and the supersymmetry breaking mass terms of the flavor quarks has to be stabilised without leading to a supersymmetry restoring runaway. Specifically, the F-terms vanish in the limit of infinite volume of the instanton. In the presence of massive $U(1)$ factors, also the D-term conditions must be taken into account.  While in the simplest classes of models this can kill the viability of the construction as stressed in  \cite{Dudas:2008qf}, one can in principle consider adding extra $U(1)$ charged matter so as to relax the D-term conditions.  Effects of D-terms for stringy embeddings of ISS type dynamics have been considered e.g. in \cite{Krippendorf:2009zz}.

A second concern is that the closed string moduli sector will often introduce additional sources of supersymmetry breaking of size comparable to or even dominating over the contributions from the gauge dynamics. Even for a supersymmetric closed moduli sector the dominance of gauge mediation over gravity mediation might be hard to ensure. We point out, though, that even in such scenarios our construction can be useful to guarantee a viable Higgs sector. Namely, in presence of extra $U(1)$ symmetries the Giudice-Masiero mechanism might fail to generate a $\mu$ term also in models of gravity mediation. E.g. K\"ahler potential terms of the form
\beq
\int d^4 x \, d^4  \theta   \, \frac{\lambda}{M}  S^{\dagger} H_u \, H_d 
\eeq
with $S$ the \emph{closed} string field whose F-term breaks supersymmetry are obviously forbidden once the operator $H_u H_d$ carries charge under some global $U(1)$ symmetry. As discussed, such symmetries are not only generic in string theory, but often also required to guarantee absence of a tree-level $\mu$ term. 
Of course in models with an F-term only due to gauge dynamics, but of comparable or competing gravity mediation contributions, our mechanism will still yield a $\mu$ term of the correct order.

\section{Discussion}

In this paper, we have presented a qualitatively new approach to solving the $\mu$ problem.  In these models, the $\mu$ term is generated directly by a non-perturbative effect in such a way that the very same non-perturbative effect generates dimensionful parameters in the SUSY breaking sector controlling the size of the F-term.  For some simple models of gauge mediation, the scale of the $\mu$ term is of the same order of magnitude as the soft masses of the MSSM.

There are two major advantages of this approach.  Firstly, because $\mu$ is generated supersymmetrically, there is no associated $B \mu$ production.  $B \mu$ could then be generated by MSSM RG flow.  One could also couple the messengers to the Higgs sector to generate $B \mu$.  Because $\mu$ has additional contributions, one can avoid the usual $B\mu$ problem of gauge mediation.  The second advantage is that one is able to generate $\mu$ in a controlled way without adding additional fields beyond those needed for SUSY breaking and mediation.  One is simply getting more out of a non-perturbative effect than just retrofitting the SUSY breaking sector.

This model is not without its share of problems.  First of all, it is clear that only special models have the right structure to even get the order of magnitude matching between $\mu$ and the gaugino masses.  In many models, even if the non-perturbative effect contributes to $\mu$ and the SUSY breaking F-term, one is many orders of magnitude from an acceptable value of $\mu$.  Eventually, in string models the moduli fields controlling the actual size of the couplings will have to be stabilized dynamically in the right regime.

Secondly, one must show that the Higgs sector is viable when including all order one factors.  As is well known from the $\mu$ / $B \mu$ problem in gauge mediation, factors of $16 \pi^2$ can ruin a perfectly good model.  In this case, there are many (so far) incalculable factors that will play an important role in determining the success of a given model.

\subsection*{Acknowledgements}
We are grateful to R. Blumenhagen, N. Craig, M. Cveti{\v c}, T. Grimm, S. Kachru, R. Richter and E. Silverstein  for discussions. Special thanks to R. Richter and E. Silverstein for comments on a draft. TW thanks the KITP Santa Barbara and the University of Pennsylvania for hospitality during parts of this work.
DG is supported in part by NSERC, the DOE under contract DE-AC03-76SF00515 and the NSF under contract 9870115. TW is supported by the DOE under contract DE-AC03-76SF00515.

\appendix

\section{Details of Stringy Realization of the Quiver}

In this appendix we provide the details of a string theoretic realisation of the quiver described in section \ref{sec_quiver}.
We engineer an $SU(5)_a \times U(1)_b \times SU(3)_c$ gauge theory in the context of Type IIB orientifolds with intersecting D7 branes wrapping holomorphic four-cycles (divisors) of a Calabi-Yau threefold $X$. We choose $X$ to be given by the manifold $Q^{(dP_9)^4}$. It arises, loosely speaking, from the quintic ${\mathbb P}_{1,1,1,1,1}[5]$ after fixing some of the complex structure moduli and replacing the resulting singular loci by four $dP_9$ surfaces. These so-called del Pezzo transitions were worked out in the present context in \cite{Grimm:2008ed, Blumenhagen:2008zz}. All geometric details relevant for our discussion can be found in sections 7.1 and 7.3 of \cite{Blumenhagen:2008zz}, whose notation we adopt. There the manifold $X$ was used to construct phenomenologically attractive $SU(5)$ GUT models. It turns out that the hidden sector of these models is just of the right form for the purpose of this article. However, for simplicity our "Standard Model" is a pure $SU(5)$ theory. The simultaneous implementation of the SUSY breaking quiver and a more realistic visible sector as in \cite{Blumenhagen:2008zz} requires more complicated geometries.

The manifold $X$ has $h^{1,1}=1+4$ and thus $5$ independent holomorphic divisors $D_5$ and $D_6, D_7, D_8, D_9$. The latter four correspond to the $dP_9$ surfaces referred to above. On a complex threefold three divisors intersect in a point, and the intersection pattern is encoded in the intersection form
\bea
\label{intform}
  I_3 &=& D_5 (D_6 D_8 + D_8 D_9 + D_6 D_9 + D_6 D_7
  + D_7 D_8 + D_7 D_9)  \nonumber \\
  && - D_6^2(D_7 + D_8 +D_9) 
  - D_7^2(D_6 + D_8 + D_9) - D_8^2(D_6 + D_7 + D_8)\\
  && - D_9^2(D_6 + D_7 + D_8 )-  D_5^3 \nonumber .
\eea
To specify the orientifold we must define a holomorphic involution $\sigma$ acting as a ${\mathbb Z}_2$ symmetry on $X$. The theory we consider is compactification of Type IIB theory on $X$ modded out by $\Omega \, (-1)^{F_L} \, \sigma$, where $\Omega$ is world-sheet parity and $F_L$ denotes the left-moving fermion number. We pick the same involution $\sigma$ as in section 7.1 of \cite{Blumenhagen:2008zz}.  It interchanges the divisors $D_7$ and $D_8$, but leaves $D_5, D_6, D_9$ invariant as a divisor. The fixed point locus was determined in \cite{Blumenhagen:2008zz} as 
\beq
D_{O7} = D_5 + D_7 + D_8.
\eeq
It is the cycle wrapped by an O7-plane.

To cancel the charge $8 D_{O7}$ of this O7-plane we introduce stacks of $N_i$ D7 branes along suitable combinations of divisors $D_i$, together with their orientifold image along the divisors $D_i'=\sigma D_i$. The rules for doing this in a consistent manner are described e.g. in section 2 of \cite{Blumenhagen:2008zz}. 

Consider the configuration

\begin{align}
\label{divdata}
    & U(5)_a:  &&  D_a = D_7,  \qquad   && D'_a = D_8,  \nonumber  \\
    & U(1)_b: &&  D_b = D_5,      \qquad   &&  D'_b = D_5,  \\
    & U(3)_c: && D_c = D_5+D_7,  \qquad    &&  D'_c = D_5+D_8, \nonumber
 \end{align}
with $N_a=5, N_b=1, N_c=3$. It cancels the O7-tadpole and realises an $SU(5)_a$ theory on $D_a$.\footnote{The diagonal $U(1)$s of $U(5)_a$ and $U(3)_c$ acquire string scale mass via the St\"uckelberg mechanism and survive merely as global symmetries.} Together with the  $U(1)_b \times SU(3)_c$ sector this is exactly as in the quiver we are aiming for.

Charged matter is localized at the intersection curves of the different divisors.
To realize the desired spectrum of four quark flavors we introduce line bundles $L_a, L_b, L_c$ (and their orientifold images) on the three brane stacks. Here we merely specify the part of the line bundles that can be written as the pullback from bundles defined on the manifold $X$, and parametrise their first Chern class as
\beq
\label{bundle_gen}
c_1(L_a) = \sum_{i} a_i D_{i+4}, \quad c_1(L_b) = \sum_{i} b_i D_{i+4}, \quad c_1(L_c) = \sum_{i} c_i D_{i+4}.
\eeq
The chirality of the massless matter from open strings stretching between the various branes is computed by the index
\beq
\label{chirality_form}
I_{ij} = - \int_X D_i \wedge D_j \wedge (c_1(L_i) - c_1(L_j)).
\eeq
If $I_{ij} >0 $ there is an excess of chiral multiplets in the bifundamental representation $({ \overline{\fund}}_i, \fund_j)$. Similar  expressions exist for matter in the (anti-)symmetric representations \cite{Blumenhagen:2008zz}.
If $D_i =D_j$, the vector-like matter content has to be determined by a genuine cohomology computation.
With the help of the intersection form (\ref{intform}) we collect the formulas relevant for the chiral spectrum in table \ref{tab_chir_gen}.

\begin{table}[htbp] 
\renewcommand{\arraystretch}{1.5} 
\begin{center} 
\begin{tabular}{|c||c|c|c|} 
\hline 
\hline 
chirality & $U(5)_a$ & $U(1)_b$ & $U(3)_c$  \\ 
\hline \hline 
$2 (a_1 - a_3 - a_4)$ & $\Yasymm_{\,(2)}$ & $1$ & $1$ \\ 
$-a_2 - a_4 - a_5 + b_2 + b_4 + b_5$                         &  $\antifund_{\, (-1)}$ & $\fund_{\, (1)}$  & $1$ \\ 
$ a_2 + a_4 + a_5 + b_2 + b_3 + b_5$ & $\fund_{\, (1)}$ & $\fund_{\, (1)}$  & $1$ \\ 
$-2 b_2 - b_3 - b_4 - 2 b_5$ & $1$   & $\Ysymm_{\,(2)}$ &  $1$ \\ 
$-2 b_1 + 2 b_2 + b_3 + b_4 + 2b_5$ & $1$   & $\Yasymm_{\,(2)}$ &  $1$ \\ 
$ b_1 - b_2 - b_4 - b_5 - c_1 + c_2 + c_4 + c_5  $ & $1$ &    $\antifund_{\, (-1)}$ & $\fund_{\, (1)}$  \\
$-b_1 + b_2 + b_3 + b_5 - c_1 + c_2 + c_4 + c_5  $ & $1$ &    $\fund_{\, (1)}$ & $\fund_{\, (1)}$  \\
$ 4 c_2 + 4 c_5   $  &  $1$ & $1$ & $\Yasymm_{\,(2)}$  \\ 
\hline 
\end{tabular} 
\caption{Non-vanishing chiral spectrum for intersecting D7 brane model. The subscripts give 
 the $U(1)$ charges.} 
\label{tab_chir_gen} 
\end{center} 
\end{table} 

\subsubsection*{The model}

To define a concrete string vacuum we must specify the background value of the NS $B$-field which enters the quantisation condition of the line bundles (see \cite{Blumenhagen:2008zz} for details). In this paper we consider configurations with 
\beq
\label{B-field}
B = \frac{1}{2} D_5.
\eeq
In view of the non-spin property of the divisors $D_a, D_b, D_c$ and the intersection form (\ref{intform}) one finds that with this choice all coefficients in (\ref{bundle_gen}) have to be integer.

Consider now the special configuration
\bea
\label{bundata}
&& c_1(L_a) = 8 D_5 + 8 D_7, \quad c_1(L_b) = 5 D_6 - D_7 -D_8, \nonumber \\
&& c_1(L_c) = D_5 + D_6 + D_8 - D_9.
\eea
We can read off the chiral spectrum from table \ref{tab_chir_gen}. 
From $I_{b'c} = 4= I_{cb}$ we find the desired four pairs of quarks $q_i, \tilde q_j$ in the representation $({\bf 3}_c,1_b)$ and $({\bf \ov 3}_c,1_b)$. In this model it so happens that $I_{a'b} = 4 =I_{ab}$ account for four pairs of fields Higgs doublets $(H_u, H_d)$ of charge $({\bf 5}_a, 1_b)$ and $({\bf \ov 5}_a, 1_b)$, respectively. 
On the other hand, there are generically no massless fields in the $a-c$ sector due to $I_{ac}=0=I_{a'c}$.

\subsubsection*{Consistency conditions and D-term constraints}
Apart from cancellation of D7 brane charge various consistency conditions have to be met for the model to represent a viable string construction.
One can show that for the above choice of bundles a potential D5 brane tadpole is cancelled. To analyze the K-theory charge constraints one has to verify if for each probe brane carrying a symplectic gauge group the index of states in the fundamental representation is even \cite{Uranga:2000xp}. With the choice (\ref{B-field}) $D_5$, $D_6$ and $D_9$ can carry trivial line bundles and can thus be invariant under the orientifold.  While for $D_5, D_6$ the probe brane criterion is satisfied, it is violated for $D_9$. Since it is hard to determine if this divisor indeed carries symplectic (as opposed to orthogonal) Chan-Paton factors, we do present this model as a promising example, but not without pointing out the potential subtlety in connection with K-theory charge cancellation. 
On the other hand, the model does slightly overshoot the D3 brane tadpole \cite{Blumenhagen:2008zz}
\bea
N_{D3} + N_{gauge} = 10,
\eea
since the flux-induced D3-charge on the D7 branes is $N_{gauge} = - \frac{1}{2} \sum_i N_i \, c_1^2(L_i) = 15$.
This overshooting is rooted in the simplistic form of the chosen orientifold projection, whose fixed-point locus exhibits only a relatively small Euler number. Cancellation of the D3-tadpole e.g. by anti D3 branes would break supersymmetry at the string scale and clearly undermine the purpose of the quiver.
We trust that these technical obstacles can be overcome in more sophisticated geometries.
We also note that solutions leading to a single (as opposed to four or more) Higgs pairs can be found, but at the price of inducing half-integer  $N_{gauge}$. Since it is not clear how to cancel the D3 brane tadpole in this case even in principle, irrespective of the question of overshooting, we do not present these examples here.

Finally, the D-term supersymmetry conditions allow for a solution on the boundary of the K\"ahler cone for non-zero Vevs of all charged matter fields. 
We leave it to the reader to convince themselves, from the analysis in section 7.1 of \cite{Blumenhagen:2008zz}, that the Fayet-Iliopoulos terms vanish for a K\"ahler form $J=r_i K_i$ with 
\bea
r_2 =  r_3 = x, \quad r_4 = \frac{1}{4} x.
\eea

\subsubsection*{Instanton sector}

We now show that the model possesses an instanton sector which simultaneously generates a $\mu$ term and a mass for the quarks. 
Consider a Euclidean D3 brane wrapping the cycle $D_6$. With the choice of  B-field $(\ref{B-field})$ a brane along $D_6$ carrying trivial gauge flux satisfies the Freed-Witten quantisation condition. Such a configuration is invariant under the orientifold projection and thus yields either orthogonal or symplectic gauge group. Since $D_6$ is transverse to the O7-plane a spacetime-filling D-brane along $D_6$ carries symplectic Chan-Paton factors. By general arguments (see e.g. \cite{Blumenhagen:2009qh} for a review) a D3 instanton along $D_6$ is therefore of $O(1)$ type. This means it has the right universal zero mode structure $d^4 x d^2 \theta$ to contribute to the superpotential. Furthermore $D_6$ is a rigid $dP_9$ so that the instanton carries no extra deformation modes that would spoil the superpotential contribution.

The charged zero modes are computed from formula (\ref{chirality_form}) taking into account that the instanton $\cal E$ on $D_6$ carries trivial line bundle. 
From $I_{b{\cal E}} =2$ one finds precisely two charged zero modes $\alpha, \beta$ in the representation $(-1_b,1_{\cal E})$ (where $1_{\cal E}$ refers to the $O(1)$ symmetry of the instanton).  
Note that open strings in the $b-\cal E$ and in the ${\cal E} - b'$ sector are identified for ${\cal E} = {\cal E}'$ so that we can think of these modes as $\alpha_{{\cal E} b'}$ and $\beta_{b \cal E}$.
It is then clear that by gauge invariance the instanton effective action can contain the four-point couplings
\beq
\label{inst_coup2}
S_{int} = c^{i j}_F \, M_s^{-2} \, \alpha \,  q_i \, \tilde{q_j} \, \beta + c_{\mu} \, M_s^{-2} \, \alpha \, H_u \, H_d \, \beta.
\eeq
For these couplings to really exist, the matter curves of the involved states must meet in a point so that the wave-functions have non-zero overlap. This indeed follows from a closer look at the geometry: The modes
$\alpha$ and $\beta$ are localised on the curve $C_{56} = D_5 \cap D_6$, while $H_u, H_d$ are localised on $C_{57}= D_5 \cap D_7$. Both of these loci are elliptic curves; namely they are the elliptic fibers of the $dP_9$ surfaces $D_6$ and $D_7$, respectively. We know from (\ref{intform}) that $D_5, D_6, D_7$ intersect in one point, which has to be the point common to $C_{56}$ and $C_{57}$. Together with a similar reasoning for $D_5, D_7$ and the divisor $D_c= D_5+D_7$ this establishes the existence of both terms in (\ref{inst_coup2}) required for the non-perturbative generation of a $\mu$ term and quark masses.
Clearly, the actual computation of the moduli dependence of $c_F$ and $c_{\mu}$ is much more involved and beyond our scope here.

\subsubsection*{Messenger sector}

The divisor $D_c=D_5+D_7$ can be understood as the recombination of the two sparate divisors $D_{c_1}=D_5$ and $D_{c_2}=D_7$ into a single object. 
The recombination moduli are given by the open strings in the $D_5-D_7$ sector localised on the elliptic intersection curve $C_{57} = D_5 \cap D_7$. For zero gauge flux on $D_5$ and $D_7$ they are counted by the cohomology groups $H^i(C_{57}, {\cal O}) = (1,1)$. Here it is crucial that the intersection curve is indeed a $T^2$ to allow for a vector-like pair of states $\phi_+$ and $\phi_-$ in the $D7-D5$ and $D5-D7$ sector, respectively, corresponding to the elements of $H^0(T^2, \cal O)$ and $H^1(T^2, \cal O)$. Since only one combination of $\phi_+, \phi_-$ is constrained by the D-terms (\ref{Dterms}), we are left with an open string modulus in the manner explained in section \ref{sec_quiver}.
Note that the curvature induced D3-charges of $D5$ and $D7$ add up correctly, in agreement with charge conservation upon brane recombintation. This is because the Euler characters $\chi(D5)=13$ and $\chi(D7)=12$ computed in section 7.1 of \cite{Blumenhagen:2008zz} sum up to $\chi(D5+D7) = 25$.

The presence of a line bundle $L_c$ on $D_c$ does not affect this analysis much.
The system $(D_5+D_7, L_c|_{D_5 + D_7})$ splits into $(D_5, L_c|_{D_5})$ and $(D_7, L_c|_{D_7})$, again
in agreement with charge conservation on the D-branes.\footnote{Specifically, also the flux induced D3-charge $\int_X (D5+D7) \wedge c_1(L_c)^2$ is obviously conserved by linearity of the intersection form.} Since we  still find $H^i(C_{57}, L_c \otimes L_c^*) = H^i(C_{57}, {\cal O}) = (1,1)$ on the split locus, the gauge flux $L_c$ does not lift the modulus associated with one of the recombination fields.

The massive spectrum of the recombined brane can be inferred from the massless states on the split locus given by $\phi_+ = 0 = \phi_-$. In the presence of superpotential couplings to $\phi_{\pm}$ these fields acquire open string modulus dependent masses away from this locus. 
This means that there will be one massive pair of messengers $\Psi, \tilde \Psi$ for non-zero vev of the recombination modulus $\phi_-$ if the system at $\phi_{\pm} =0$ exhibits precisely one massless pair $\Psi, \tilde \Psi$ with superpotential coupling $\phi_- \Psi \tilde \Psi$, as discussed in section \ref{sec_quiver}. In our model
\beq
\label{psi-chir}
I_{c_2, a} = - \int D_7 \wedge D_7 \wedge (c_1(L_c) - c_1(L_a)) = 1
\eeq
indeed shows the existence of one messenger $\Psi$ in the ${\bf 5}$ of $SU(5)_a$.
Similarly $I_{c_1,a} = - \int D_5 \wedge D_7 \wedge (c_1(L_c) - c_1(L_a))  = -1$ corresponds to $\tilde \Psi$.
Note that the analagous intersections with $(D_a,L_a)$ replaced by the orientifold image $(D_a',L_a')$ vanish.
Since $D5$ and $D7$ intersect along a curve, there will appear no extra vector-like states in the $(D_5,L_c) - (D_7,L_a)$ sector. For the $c_2 - a$ sector, by contrast, (\ref{psi-chir}) merely counts the chiral index at $\phi_{\pm} =0$, and one really has to compute the full cohomology groups to detect extra vector-like pairs. If present, these will lead to more than just one pair of massive messengers for $\phi_- \neq 0$. We do not perform this analysis here, but note that such extra pairs can generically be removed by twisting the line bundles with bundles in the relative cohomology of the divisors (see \cite{Blumenhagen:2008zz} for details). This is at the cost of introducing extra D3-charge, thus worsening the overshooting. The same remark holds for the sectors $D_{c_2 }- D_b$ and $D_{c_1} - D_b$, which would lead, if non-trivial, to massive quarks of potentially the same mass scale as the messengers.

\clearpage
\bibliography{muproblem}
\bibliographystyle{utphys}

\end{document}